\documentclass[apj]{emulateapj}

\usepackage{apjfonts}
\usepackage{epsfig}
\usepackage{graphicx}
\usepackage{url}
\usepackage{natbib}
\usepackage{float}
\usepackage{amsmath}
\usepackage[bf, sf, FIGTOPCAP, nooneline, tight]{subfigure}
\usepackage{graphicx}
\usepackage{dcolumn}
\usepackage{bm}
\usepackage[usenames,dvipsnames]{color}
\usepackage[citebordercolor={0 1 0}]{hyperref}
\usepackage{amssymb}
\shorttitle{Probing Cosmic Acceleration}
\shortauthors{Luo X.L. et al.}

\newcommand{\be}{\begin{equation}}
\newcommand{\ee}{\end{equation}}
\newcommand{\bq}{\begin{eqnarray}}
\newcommand{\eq}{\end{eqnarray}}

\def\({\left(}
\def\){\right)}

\begin{document}

\title{Probing Cosmic Acceleration by Using Model-Independent Parametrizations and Three Kinds of Supernova Statistics Techniques}

\author{Xiaolin Luo\altaffilmark{1},
Shuang Wang\altaffilmark{1},
Sixiang Wen\altaffilmark{1}
}

\email{luoxl23@mail2.sysu.edu.cn}
\email{wangshuang@mail.sysu.edu.cn (Corresponding author)}
\email{wensx@mail2.sysu.edu.cn}

\altaffiltext{1}{School of Physics and Astronomy, Sun Yat-Sen University, Guangzhou 510297, P. R. China}

%\date{\today}

\begin{abstract}
  In this work, we explore the evolution of the dark energy equation of state $\omega$ by using Chevalliear-Polarski-Linder (CPL) parametrization and the binned parametrizations. For binned parametrizations, we adopt three methods to choose the redshift interval: I. Ensure that  ``$\triangle z=const$'', where $\triangle z$ is the width of each bin; II. Ensure that ``$n\triangle z=const$'', where n is the number of SNIa in each bin; III. Treat redshift discontinuity points as models parameters, i.e. ``free $\triangle z$''. For observational data, we adopt JLA type Ia supernova (SNIa) samples, SDSS DR12 data, and Planck 2015 distance priors. In particular, for JLA SNIa samples, we consider three statistic techniques: I. Magnitude statistics, which is the traditional method; II. Flux statistics, which reduces the systematic uncertainties of SNIa; III. Improve flux statistics, which can reduce the systematic uncertainties and give tighter constrains at the same time. The results are as follows: (1) For all the cases, $\omega = -1$ is always satisfied at $1\sigma$ confidence regions; It means that $\Lambda$CDM is still favored by current observations. (2) For magnitude statistics, ``free $\triangle z$'' model will give the smallest error bars; this conclusion does not hold true for flux statistics and improved flux statistic. (3) The improved flux statistic yields a largest present fractional density of matter $\Omega_m$; in addition, this technique will give a largest current deceleration parameter $q_0$ , which reveals a universe with a slowest cosmic acceleration.

\end{abstract}

\maketitle

\

\section{Introduction}\label{intro}
Since the discovery of cosmic acceleration in 1988 \citep{AGRiess1998,Perlmutter1999}, it's widely believed that there exists mysterious dark energy (DE) that drives the current accelerating expansion of the universe \citep{Padmanabhan2003,Frieman2008,MLi2011,Bamba2012,MLi2013}.  To explain this strange phenomenon, numerous theoretical models have been proposed, such as  $\Lambda CDM$ \citep{Einstein1917}, quintessence \citep{Caldwell1998,Zlatev1999}, phantom \citep{Caldwell2002,Carroll2003}, k-essence \citep{Picon1999,Chiba2000},  Chaplygin gas \citep{Kamenshchik2001,Bento2002}, holographic DE \citep{hde,MLi2009a,MLi2009b,SWang2017}, agegraphic DE \citep{rgcai2007,weicai2008}, Ricci DE \citep{Gao2009},  Yang-Mills condensate \citep{YZhang2007,SWang2008,WZ2008}.

In addition to specific DE models, another way of exploring the nature of DE is to adopt model-independent reconduction \citep{Shapiro2004,wangprl2004,Ywang2007,YunWang2009,wll2011}.
A popular approach is the so-called specific ansatz. Here we consider the most popular Chevalliear-Polarski-Linder \citep{CPL,Linder:2002et}. Another important method is the so-called binned parametrization, which was firstly proposed by Huterer and Starkman based on the principal component analysis (PCA) \citep{Huterer2003}. The basic idea of binned parametrization is dividing the redshift range into different bins and setting equation of state (EOS) $w$ as piecewise constant in redshift $z$.

It must be stressed that there are different opinions in the literature about the optimal choice of the discontinuity points of redshift. \cite{ywang2010} argued that the width of each redshift bin, i.e. $\triangle z$, should be a constant (hereafter we call it ``const $\triangle z$'' model). \cite{AGRiess2007} argued that the number of SNIa in a bin (i.e., n), times the width of this bin should be a constant (hereafter we call it  ``const $n\triangle z$'' model).  In \cite{QGHuang2009}, one of the current authors and his collaborators showed that the discontinuity points of redshift can be treated as model parameters in performing cosmology-fits (hereafter we call it ``free $ \triangle z$'' model. In this work, we will consider all the three binned models.

For observational aspect, we mainly focus on type Ia supernova (SNIa) that can be regard as cosmological standard candles to measure directly the expansion history of the Universe \citep{weinberg2013}. In the literature, people always make use of the distance-redshift relation $\mu(z)$ to calculate the $\chi^2$ function of SNIa. From now on, we call this statistic method of SNIa as ``magnitude statistic'' (MS). However, a lot of recent studies showed that the MS technique suffers from the systematic uncertainties of SNIa \citep{Hu2016}. For example, it had been proved that, in the framework of MS statistic,  SN color luminosity parameter $\beta$ should evolve along with redshift z at 5$\sigma$ confidence level \citep{WangWang2013,wlz2013,Mohlabeng2014,WWGZ2014,WWZ2014,WGHZ2015,li2016}. To overcome the shortcut of MS, \cite{Wang2000} proposed a ``flux-averaged''(FA) technique, which average the observed flux of SNIa at a series of uniformly divided redshift bins. Hereafter, we call the statistic method that based on FA technique as ``flux statistic''(FS) method, which can effectively reduce the systematic uncertainties of SNIa \citep{WangTegmark05,YunWang2009,Wang12CM}. However, adopting the FS method will yield larger error bars for various model parameter. Therefore, in 2013, one of the current authors and Wang \citep{WangWang2013} developed an improved FA method, which only use the FA technique at high redshift region. Hereafter, we call this improved FA method as  ``improve flux statistics''(IFS), which can reduce systematic uncertainties of SNIa and give tighter DE constraints at the same time \citep{Wang2017,Wen2018}. In this paper, we consider all the three SNIa analysis techniques.

In this paper, we make use of the current cosmological observations, including SNIa, baryon acoustic oscillations (BAO) and cosmic microwave background (CMB), to constrain various model-independent DE reconductions.
In particular, for a comprehensive study, we consider all the three binned models (i.e., const $\triangle z$, const $n\triangle z$ and $free \triangle z$) and all the three SNIa analysis techniques (i.e., MS, FS and IFS).
This paper is organized as follows:
In Section~\ref{models} , we introduce the theoretical models, including Chevalliear-Polarski-Linder (CPL) and three kinds of binned parametrization models.
In Section~\ref{data}, we introduce all the observational data, and  describe the details of using SNIa, BAO and CMB.
Finally, we present our results in Section~\ref{secresult} and conclude our work in Section~\ref{conclu}.

\

\section{Theoretical models}\label{models}
In a spatially flat universe, the Friedmann equation can be rewritten as
\be
\label{Fried}
H=H_0\sqrt{\Omega_{\rm{r}}(1+z)^4+\Omega_{\rm{m}}(1+z)^3+\Omega_{\rm{de}}X(z)}.
\ee
Here $\Omega_{\rm{m}}$, $\Omega_{\rm{r}}$ and $\Omega_{\rm{de}}$ are the present fractional
densities of matter, radiation and dark energy, respectively, $H_0$ is the present-day value of the Hubble parameter $H(z)$.
And the radiation density parameter $\Omega_r$ is given by \citep{WangyunWangshuang2013},
\be
\Omega_{r}=\Omega_{m}/(1+z_{\rm eq}),
\ee
where $z_{\rm eq}=2.5\times10^4\Omega_{m}h^2(T_{\rm cmb}/2.7\,{\rm K})^{-4}$, $T_{\rm cmb}=2.7255\,{\rm K}$, and $h$ is the reduced Hubble constant.
Note that $X(z)$ is given by the specific dark energy models,
\be
X(z)\equiv \rho_{de}(z)/\rho_{de}(0)=\exp[3\int_0^z\frac{1+w(z')}{1+z'}dz']
\ee
where $\omega \equiv p_{de}/\rho_{de}$ is the EOS of DE, $ p_{de}$ and $\rho_{de}$ are pressure and density of DE, respectively. In this paper, we consider two kind of DE model-independent parametrization approaches listed below.
\subsection{Chevallier-Polarski-Linder parametrization}
Firstly, we consider the Chevallier-Polarski-Linder parametrization, with the EOS parameterized as
\begin{equation}
\label{cpl1}
w(z)=w_0+w_a\frac{z}{1+z},
\end{equation}
where $w_0$ and $w_a$ are constant parameters. The corresponding $X(z)$ can be
expressed as
\begin{equation}
\begin{aligned}
 X(z)=(1+z)^{3(1+w_{\rm{0}}+w_{\rm{a}})}\exp\left(-\frac{3w_{\rm{a}}z}{1+z}\right).
 \end{aligned}
\label{xz}
\end{equation}

\subsection{Three kinds of binned parametrization}
  For the case where EOS $\omega$ is piecewise constant in redshift, we only consider the case of 3 bins.
  So we can reconstruct it as
    \be
       w(z) = \left\lbrace
       \begin{array}{ll}
        w_1 &\quad 0<z<z_1\\
        w_2 &\quad z_1<z<z_2\\
        w_3 &\quad z_2<z.
       \end{array}
    \label{wz}
    \right.
    \ee
  where $w_1,w_2,w_3$ are free parameters and will be determined by MCMC method \citep{Lewis2002}, $z_1, z_2 $ will be determined by three kinds of binned parametrization.
  In addition, the corresponding $X(z)$ takes the form
  \be
  X(z_{n-1}<z<z_n)=(1+z)^{3(1+w_n)}\prod\limits _{i=0}^{n-1}(1+z_i)^{3(w_i-w_{i+1})},
  \ee
  where $w_i$ is the EOS parameter in the $i^{th}$ redshift bin defined by an upper boundary at $z_i$.

  As mentioned above, we consider three binned methods as follow:
  \begin{itemize}

   \item
   Firstly, we choose $z_1=0.5$, $z_2=1.0$ to get the $\triangle z=0.5$. We will call it ``const $\triangle z$''.
   \item
   Secondly, we choose $z_1=0.2481$, $z_2=0.6787$ to get $n\triangle z=97.7$. It's called ``const $n\triangle z$''.
   \item
   Thirdly, we determine the values of $z_1$ and $z_2$ by performing a best-fit analysis. It must be emphasized that adopting different SNIa statistics techniques will give different $z_1$ and $z_2$ (shown in Table~\ref{frint}). we will call this case ``free $\triangle z$''.
\end{itemize}

For the convenience of readers, we list the details about the redshift discontinuity points of three binned methods in Table~\ref{frint}.

    \begin{table*}
    \caption{\label{frint} The redshift discontinuity points of three binned methods }\centering
    \begin{tabular*}{\textwidth}{@{}l*{15}{@{\extracolsep{0pt plus12pt}}l}}
    \hline\hline
    Interval chosen&Statistics method    & $z_1$   & $z_2$ \\
    \hline
    const$\triangle z$&All&0.5000&1.0000\\
    const $n\triangle z$&All&0.2481&0.6787\\
    \hline
                     &Magnitude statistics  &   0.6482    &1.2808   \\
    free$\triangle z$&Flux statistics     &  0.4489    & 1.4330  \\
                     &Improved flux statistics    &   0.1410    & 0.7376 \\
    \hline
    \end{tabular*}
    \end{table*}

\

\section{Observational data}\label{data}
In our work, we used various cosmological observations, including SNIa, BAO and CMB, to perform cosmology-fits. Therefore,
\be
\chi^2=\chi^2_{SNIa}+\chi^2_{BAO}+\chi^2_{CMB}.
\ee
In addition, we perform a MCMC likelihood analysis \citep{Lewis2002} to obtain $O(10^6)$ samples for each set of results presented in this paper.

Moreover, figure of merit (FoM) is a very useful tool to assess the ability of constraining DE of an experiment. In this paper, we adopt a generalized FoM \citep{YunWang2008} given by:
\begin{equation}
FoM=\frac{1}{\sqrt{det~Cov(f_1,f_2,f_3,\cdots)}},
\end{equation}
where $Cov(f_1,f_2,f_3,\cdots)$ is the covariance matrix of the chosen set of DE parameters.
As is well known, larger FoM indicates better accuracy.

In the following, we will describe how to calculate the $\chi^2$ functions of SNIa, BAO and CMB.

  \subsection{Type Ia supernovae}
  For the SNIa data, we make use of ``Joint Light-curve Analysis'' (JLA) dataset \citep{Betoule2014}. In particular, we introduce in details the three kinds of SNIa statistics techniques (i.e., MS, FS, and IFS).

    \subsubsection{Magnitude statistics}
Firstly, we describe MS in this section.
Theoretically, the distance modulus $\mbox{\bf $\mu$}_{th}$ in a flat universe can be written as
\be
  \mbox{\bf $\mu$}_{th} = 5 \log_{10}\bigg[\frac{d_L(z_{hel},z_{cmb})}{Mpc}\bigg] + 25,
\ee
where $z_{cmb}$ and $z_{hel}$ are the CMB restframe and heliocentric redshifts of SNIa.
The luminosity distance ${d}_L$ is given by
\be
  {d}_L(z_{hel},z_{cmb}) = (1+z_{hel}) r(z_{cmb}),
\ee

where $r(z)$ is given by
\be \label{eq:rz}
  r(z)= cH_0^{-1}\int_0^z\frac{dz'}{E(z')},
\ee
$c$ is the speed of light, and $E(z)\equiv H(z)/ H_0$.
The observation of distance modulus $\mbox{\bf $\mu$}_{obs}$ is given by an empirical linear relation:
\be
  \mbox{\bf $\mu$}_{obs}= m_{B}^{\star} - M_B + \alpha_0 \times X_1
  -\beta_0 \times {\cal C},
\ee
where $m_B^{\star}$ is the observed peak magnitude in the rest-frame
\text{of the} $B$ band,
$X_1$ describes the time stretching of light-curve, and ${\cal C}$ describes the
supernova color at maximum brightness.
Note that $\alpha_0$ and $\beta_0$ are SN stretch-luminosity parameter and SN color-luminosity parameter, respectively.
In addition, $M_B$ is the absolute B-band magnitude,
which relates to the host stellar mass $M_{stellar}$ via a simple step function

\begin{equation}
  \label{eq:mabs}
    M_B = \left\lbrace
   \begin{array}{ll}
    M^1_B &\quad \text{if}\quad  M_{stellar} < 10^{10} M_{\odot}\,,\\
    M^2_B &\quad \text{otherwise.}
    \end{array}
    \right.
\end{equation}
where $M_{\odot}$ is the mass of sun.
The $\chi^2$ of JLA data can be obtain by
\be
\label{eq:chi2_SN}
\chi^2_{SNIa} = \Delta \mbox{\bf $\mu$}^T \cdot \mbox{\bf Cov}^{-1} \cdot \Delta\mbox{\bf $\mu$},
\ee
where $\Delta \mbox{\bf $\mu$}\equiv \mbox{\bf $\mu$}_{obs}-\mbox{\bf $\mu$}_{th}$
is the data vector and $\mbox{\bf Cov}$ is the total covariance matrix, which can be calculated as
\be
\mbox{\bf Cov}=\mbox{\bf D}_{\rm stat}+\mbox{\bf C}_{\rm stat}
+\mbox{\bf C}_{\rm sys}.
\ee
Here $\mbox{\bf D}_{\rm stat}$ is the diagonal part of the statistical
uncertainty, which is given by
\begin{eqnarray}
\mbox{\bf D}_{\rm stat,ii}&=&\left[\frac{5}{z_i \ln 10}\right]^2 \sigma^2_{z,i}+
  \sigma^2_{\rm int} +\sigma^2_{\rm lensing} + \sigma^2_{m_B,i}\nonumber\\
&&+ \alpha_0^2 \sigma^2_{X_1,i}+\beta_0^2 \sigma^2_{{\cal C},i} + 2 \alpha_0 C_{m_B X_1,i} \nonumber\\
&&- 2 \beta_0 C_{m_B {\cal C},i}  -2\alpha_0\beta_0 C_{X_1 {\cal C},i},
\end{eqnarray}
where the first three terms account for the uncertainty in redshift due to peculiar velocities,
the intrinsic variation in SN magnitude and the variation of magnitudes caused by gravitational lensing.
$\sigma^2_{m_B,i}$, $\sigma^2_{X_1,i}$, and $\sigma^2_{{\cal C},i}$
denote the uncertainties of $m_B$, $X_1$ and ${\cal C}$ for the $i$-th SN.
In addition, $C_{m_B X_1,i}$, $C_{m_B {\cal C},i}$ and $C_{X_1 {\cal C},i}$
are the covariances between $m_B$, $X_1$ and ${\cal C}$ for the $i$-th SN.
Moreover, $\mbox{\bf C}_{\rm stat}$ and $\mbox{\bf C}_{\rm sys}$
are the statistical and the systematic covariance matrices, given by
\begin{equation}
\mbox{\bf C}_{\rm stat}+\mbox{\bf C}_{\rm sys}=V_0+\alpha_0^2 V_a + \beta_0^2 V_b +
2 \alpha_0 V_{0a} -2 \beta_0 V_{0b} - 2 \alpha_0\beta_0 V_{ab},
\end{equation}
where $V_0$, $V_{a}$, $V_{b}$, $V_{0a}$, $V_{0b}$ and $V_{ab}$ are six matrices which will be given in \cite{Betoule2014}.
    \subsubsection{Flux statistics}
FS based on the FA technique, which is very useful to reduce the systematic uncertainties of SNIa \citep{Wang2004,WangTegmark05,YunWang2009}.
The original FA method divide the whole redshift region of SNIa into a lot of bins,
where the redshift interval of each bin is $\delta z$.
In the following, we will introduce the specific steps of FA \citep{Wang12CM}:

(1) Convert the distance modulus of SNIa into ``fluxes'',
\be
\label{eq:flux}
F(z_l) \equiv 10^{-(\mu_0^{\rm obs}(z_l)-25)/2.5} =
\left( \frac{d_L^{\rm obs}(z_l)} {\mbox{Mpc}} \right)^{-2}.
\ee
Here $z_l$ represent the CMB restframe redshift of SN.

(2) For a given set of cosmological parameters $\{ {\bf s} \}$,
calculate ``absolute luminosities'', \{${\cal L}(z_l)$\},
\be
\label{eq:lum}
{\cal L}(z_l) \equiv d_L^2(z_l |{\bf s})\,F(z_l).
\ee

(3) Flux-average the ``absolute luminosities'' \{${\cal L}^i_l$\}
in each redshift bin $i$ to obtain $\left\{\overline{\cal L}^i\right\}$:
\be
 \overline{\cal L}^i = \frac{1}{N_i}
 \sum_{l=1}^{N_i} {\cal L}^i_l(z^{(i)}_l),
 \hskip 1cm
 \overline{z_i} = \frac{1}{N_i}
 \sum_{l=1}^{N_i} z^{(i)}_l.
\ee

(4) Place $\overline{\cal L}^i$ at the mean redshift $\overline{z}_i$ of
the $i$-th redshift bin, now the binned flux is
\be
\overline{F}(\overline{z}_i) = \overline{\cal L}^i /
d_L^2(\overline{z}_i|\mbox{\bf s}).
\ee
with the corresponding flux-averaged distance modulus:
\be
\overline\mu^{obs}(\overline{z}_i) =-2.5\log_{10}\overline{F}(\overline{z}_i)+25.
\ee

(5) Calculate the covariance matrix of $\overline{\mu}(\overline{z}_i)$
and $\overline{\mu}(\overline{z}_j)$:
\begin{equation}
\begin{aligned}
\mbox{Cov}&\left[\overline{\mu}(\overline{z}_i),\overline{\mu}(\overline{z}_j)\right]
=\frac{1}{N_i N_j \overline{\cal L}^i \overline{\cal L}^j}\\
 &\sum_{l=1}^{N_i} \sum_{m=1}^{N_j} {\cal L}(z_l^{(i)})
{\cal L}(z_m^{(j)}) \langle \Delta \mu_0^{\rm obs}(z_l^{(i)})\Delta
\mu_0^{\rm obs}(z_m^{(j)})
\rangle
\end{aligned}
\end{equation}
where $\langle \Delta \mu_0^{\rm obs}(z_l^{(i)})\Delta \mu_0^{\rm obs}(z_m^{(j)})\rangle $
is the covariance of the measured distance moduli of the $l$-th SNIa
in the $i$-th redshift bin, and the $m$-th SNIa in the $j$-th
redshift bin. ${\cal L}(z)$ is defined by Eqs.(\ref{eq:flux}) and (\ref{eq:lum}).

(6) For the flux-averaged data, $\left\{\overline{\mu}(\overline{z}_i)\right\}$,
calculate
\be
\label{eq:chi2_SN_fluxavg}
\chi^2_{SNIa} = \sum_{ij} \Delta\overline{\mu}(\overline{z}_i) \,
\mbox{Cov}^{-1}\left[\overline{\mu}(\overline{z}_i),\overline{\mu}(\overline{z}_j)
\right] \,\Delta\overline{\mu}(\overline{z}_j)
\ee
where
\be
\Delta\overline{\mu}(\overline{z}_i) \equiv
\overline{\mu}^{obs}(\overline{z}_i) - \mu^p(\overline{z}_i|\mbox{\bf s}),
\ee
and
\be
\overline\mu^p(\overline{z}_i) =-2.5\log_{10} F^p(\overline{z}_i)+25.
\ee
with $F^p(\overline{z}_i|\mbox{\bf s})=
\left( d_L(\overline{z}_i|\mbox{\bf s}) /\mbox{Mpc} \right)^{-2}$.
    \subsubsection{Improved flux statistics}
As mentioned above, the improved FA method  \citep{WangWang2013} introduces a new quantity: the redshift cut-off $z_{cut}$.
For the SN samples at $z < z_{cut}$, the $\chi^2$ is computed by using the usual MS (i.e., Eq. \ref{eq:chi2_SN});
for the SN samples at $z \geq z_{cut}$, the $\chi^2$ is computed by using the ``flux statistics'' (i.e., Eq. \ref{eq:chi2_SN_fluxavg}).
Therefore, the total $\chi^2$ can be written as
\be
    \chi^2 = \left\lbrace
   \begin{array}{ll}
    \chi^2_{MS} &\quad \text{if}\quad  z < z_{cut}\,,\\
    \chi^2_{FS} &\quad \text{otherwise.}
    \end{array}
    \right.
\ee

This new method includes the advantages of MS and FS, and thus can reduce systematic uncertainties and give tighter DE constraints at the same time.

In previous works, \cite{Wang2015} applied this improved FA method to explore the JLA data,
and found that it can give tighter constraints on DE.
But in \cite{Wang2015}, only one kind of FA recipe, $(z_{cut} = 0.5, \delta z=0.04)$, was considered.
In a recent paper \citep{Wang2017}, we scanned the whole $(z_{cut}, \delta z)$ plane,
and found that adopting the FA recipe, $(z_{cut} = 0.6, \delta z=0.06)$, yielded the tightest DE constraints.
So in this paper, we will use the IFS technique with the best FA recipe $(z_{cut} = 0.6, \delta z=0.06)$.

The details of these three statistic methods of SNIa  are listed in Table \ref{SN}.

\begin{table*}
\caption{\label{SN} Details of three statistics methods of SNIa data}\centering
\begin{tabular*}{\textwidth}{@{}l*{15}{@{\extracolsep{0pt plus12pt}}l}}
\hline\hline
Statistics method    & Abbreviation   & FA recipe &   Number of SNIa samples \\
\hline
Magnitude statistics  &   MS    &N/A &  740 \\
Flux statistics     &  FS    & $z_{cut} = 0.0, \delta z=0.06$ &  21 \\
Improved flux statistics    &   IFS    & $z_{cut} = 0.6, \delta z=0.06$ &   606 \\
\hline
\end{tabular*}
\end{table*}

\subsection{Other observational data}
  \subsubsection{Baryon acoustic oscillations}
The BAO matter clustering provides a ``standard ruler'' for length scale in cosmology.
And the signals can be used to measure the Hubble parameter $H(z)$ and angular diameter distance
$D_A(z)=r(z)/(1+z)$ in the radial and tangential directions, respectively. Here the $r(z)$ is given by Eq.\ref{eq:rz}
In this paper we use the data of BOSS DR12 \citep{Alam2016},
which includes the combinations
$H(z)r_s(z_d)/r_{s,fid}$ and $D_M(z)r_{s,fid}/r_s(z_d)$.
Here $r_{s,fid} = 147.78$Mpc is the sound horizon of the fiducial model, and $D_M(z)=(1+z)D_A(z)$ is the comoving angular diameter distance.
$r_s(z_d)$ is the sound horizon at the drag epoch $z_d$, given by
\be
\label{eq:rd}
r_s(z_d)=\int_{z_d}^{\infty}\frac{c_s(z)}{H(z)}dz,
\ee
where $c_s(z)= 3^{-1/2}c[1+\frac{3}{4}\rho_{b}(z)/\rho_{r}(z)]^{-1/2}$ is the sound speed in the photon-baryon fluid.
In \cite{Alam2016}, $r_s(z_d)$ is approximated by \cite{Aubourg2015},
\begin{equation}
r_s(z_d) =
 \frac{55.154exp[-72.3(\omega_v+0.0006)^2]}{\omega_{b}^{0.12807}\omega_{cb}^{0.25351}}Mpc,
\label{eq:zd}
\end{equation}
where $\omega_v = 0.0107(\sum m_v/1.0$eV) is the density parameter of neutrinos;
$\omega_{b} = \Omega_bh^2$ is the density parameter of baryons, and $\omega_{cb}=\Omega_mh^2 - \omega_v $ is the density parameters of baryons and (cold) dark matter.
Following the \cite{Aubourg2015}, we set $\sum m_v=0.06$ for all the models we considered.

There are 6 BAO data points given in Table $7$ of \cite{Alam2016}:
\bq
& p_{1}=D_M(0.38)r_{s,fid}/r_s(z_d), &p_{1}^{data}=  1512,\nonumber\\
& p_{2}= H(0.38)r_s(z_d)/r_{s,fid}, &p_{2}^{data}= 81.2,\nonumber\\
& p_{3}=D_M(0.51)r_{s,fid}/r_s(z_d), &p_{3}^{data}= 1975, \nonumber\\
& p_{4}=H(0.51)r_s(z_d)/r_{s,fid}, &p_{4}^{data}= 90.9,\nonumber\\
& p_{5}=D_M(0.61)r_{s,fid}/r_s(z_d), &p_{5}^{data}=2307,\nonumber\\
& p_{6}=H(0.61)r_s(z_d)/r_{s,fid}, &p_{6}^{data}= 99.0.
\eq
Therefore, the $\chi^2$ function for current BAO data can be expressed as
\be
\label{eq:chi2bao}
\chi^2_{BAO}=\Delta p_{i} \left[ {\rm Cov}^{-1}_{BAO}(p_{i},p_{j})\right]
\Delta p_{j},
\hskip .5cm
\Delta p_{icur}= p_{i} - p_{i}^{data}.
\ee
The covariance matrix ${\rm Cov}_{BAO}$ is given by the on-line files of \cite{Alam2016}.

  \subsubsection{Cosmic microwave background }
CMB gives us the comoving distance to the photon-decoupling surface $r(z_*)$ and the comoving sound horizon at photon-decoupling epoch $r_s(z_*)$.
In this
paper, we use the distance priors data extracted from Planck 2015~\citep{Planck201514}.
This includes the ``shift parameter'' $R$, the
``acoustic scale'' $l_A$, and the
redshift of the decoupling epoch of photons $z_*$.

The shift parameter $R$ is given by ~\cite{Ywang2007}:
\be
R \equiv \sqrt{\Omega_{m} H_0^2} \,r(z_*)/c,
\ee
where $r(z_*)$ is the comoving distance given in Eq. \ref{eq:rz}. $z_*$ is the redshift of the photon
decoupling epoch estimated by \cite{Hu:1995en}:
\begin{equation}
\label{zstareq} z_*=1048[1+0.00124(\Omega_b
h^2)^{-0.738}][1+g_1(\Omega_m h^2)^{g_2}],
\end{equation}
here $\Omega_{b}$ is the present fractional density of baryon,and
\begin{equation}
g_1=\frac{0.0783(\Omega_b h^2)^{-0.238}}{1+39.5(\Omega_b
h^2)^{0.763}},\quad g_2=\frac{0.560}{1+21.1(\Omega_b h^2)^{1.81}}.
\end{equation}

The acoustic scale $l_A$ is defined as
\begin{equation}
\label{ladefeq} l_A\equiv \pi r(z_*)/r_s(z_*),
\end{equation}
where $r_s(z_*)$ is the comoving sound
horizon at $z_*$. The  $r_s(z)$ is given by
\be~\label{eq:rs}
r_s(z) = cH_0^{-1}\int_{0}^{a}\frac{da^{\prime}}{\sqrt{3(1+\overline{R_b}a^\prime){a^\prime}^4E^2(z^\prime)}},
\ee
where $\overline{R_b}=31500\Omega_{b}h^2(T_{cmb}/2.7K)^{-4}$,
These two distance priors, together with $\omega_b\equiv\Omega_bh^2$, provide an efficient summary of CMB data.

The $\chi^2$ function for the CMB distance prior data can be expressed as
\be
\label{eq:chi2CMB}
\chi^2_{CMB}=\Delta q_i \left[ \mbox{Cov}^{-1}_{CMB}(q_i,q_j)\right]
\Delta q_j,
\hskip .2cm
\Delta q_i= q_i - q_i^{data},
\ee
where $q_1=R(z_*)$, $q_2=l_a(z_*)$, and $q_3= \omega_b$.
The covariance matrix for $(q_1, q_2, q_3)$ is given by
\be
\mbox{Cov}_{CMB}(q_i,q_j)=\sigma(q_i)\, \sigma(q_j) \,\mbox{NormCov}_{CMB}(q_i,q_j),
\label{eq:CMB_cov}
\ee
where $\sigma(q_i)$ is the 1$\sigma$ error of observed quantity $q_i$,
$\mbox{NormCov}_{CMB}(q_i, q_j)$ is the corresponding normalized covariance matrix, which are listed in Table $4$ of ~\cite{Planck201514}.

The Planck 2015 data are
\bq
&& q_{1}^{data} = 1.7382\pm0.0088, \nonumber\\
&& q_{2}^{data} = 301.63\pm0.15, \nonumber\\
&& q_{3}^{data} = 0.02262\pm0.00029.
\label{eq:CMB_mean_planck}
\eq

\

\section{Results}\label{secresult}

\subsection{Probing cosmic acceleration with the Model-Independent Parametrizations}
 In this subsection, we will show the fitting results of CPL parametrization and the three binned parametrizations.
 It must be mentioned that all the SNIa analysis techniques, including MS, FS and IFS, are taken into account.

\subsubsection{CPL Parametrization}

\begin{figure*}[htbp]
      \centering

      \includegraphics[width=7cm,height=4cm]{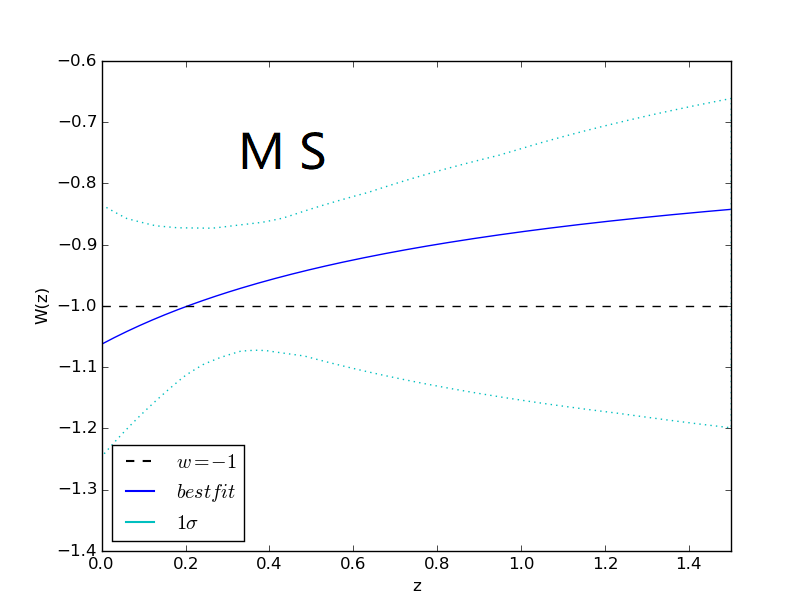}
      \includegraphics[width=7cm,height=4cm]{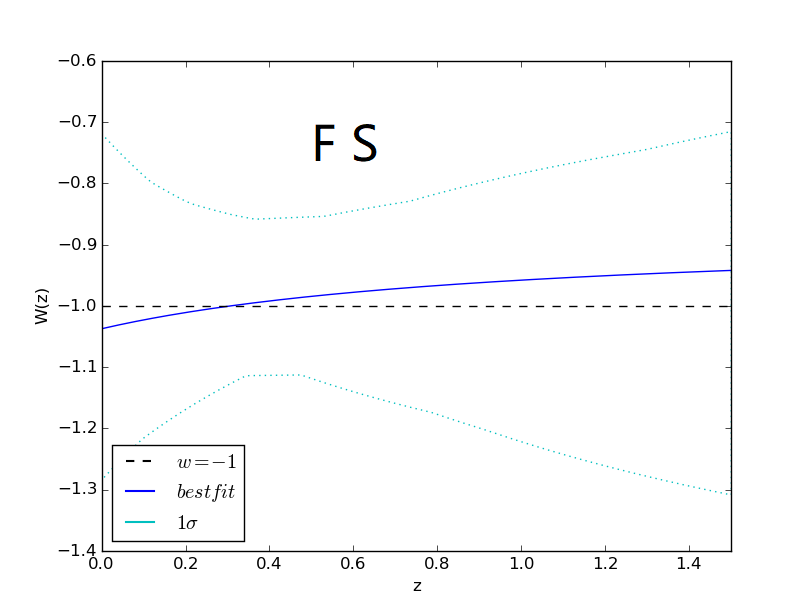}
      \includegraphics[width=7cm,height=4cm]{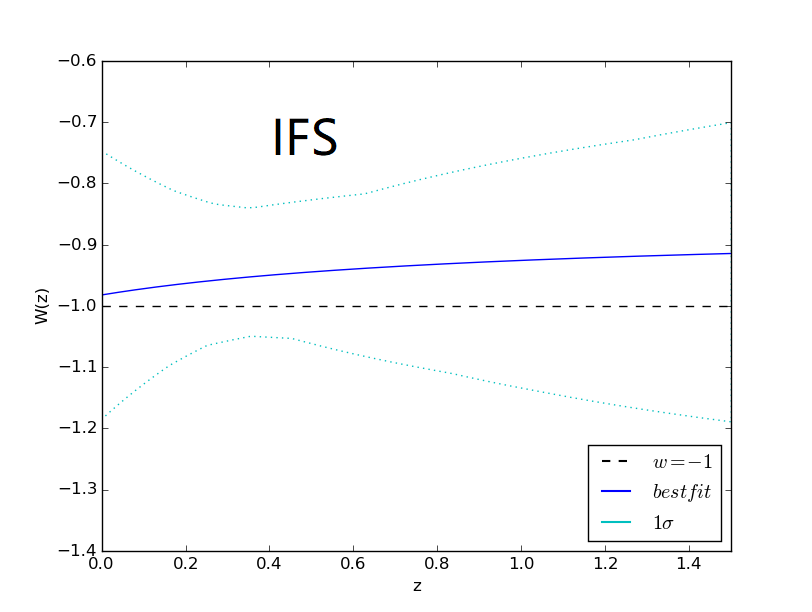}
      \caption{The evolution of w(z) for the CPL model. Three SNIa statistics techniques are used, including MS (upper left panel), FS (upper right panel) and IFS (lower panel). The solid blue lines represent the best-fit value of $\omega(z)$, the dotted cyan lines represent the 1$\sigma$ confidence region of $\omega(z)$. Moreover, the dashed black lines represent the result of $\Lambda CDM$. One can find that the result of CPL is always consistent with the result of $\Lambda CDM$ at the $1\sigma$ confidence region.}
      \label{figcpl}
    \end{figure*}

 Firstly, we consider the case of CPL parametrization.
 In Fig.\ref{figcpl}, we plot the evolution of $\omega(z)$ for the CPL parametrization. For comparing, all the three SNIa statistics techniques are used. The solid blue lines represent the best-fit value of $\omega(z)$, the dotted cyan lines represent the 1$\sigma$ confidence region of $\omega(z)$. In addition, we also present the results of $\Lambda CDM$ which are represented by dashed black lines.

 From this figure, we find that the 1$\sigma$ confidence region of  CPL's $\omega(z)$  is  consistent with the result of $\Lambda CDM$. More importantly, this conclusion is insensitive to the SNIa statistics techniques.

 Moreover, this conclusion also holds true for the cases of the SNLS3\citep{xiaodongli2011}, Union2\citep{Holsclaw2010}, Union2.1\citep{shi2012} and Constitution\citep{wei2010}. This implies that this conclusion comes into existence for all the SNIa data.

\subsubsection{Three Binned Parametrizations}

 Then, we will show the results of three binned parametrizations.
 For a more systematic and more comprehensive study, all the SNIa statistics techniques are taken into account. Moreover, we consider three kinds of binned models, i.e., ``const $\triangle z$'', ``const n$\triangle z$'' and ``free $\triangle z$''. Different binned modles have different redshift discontinuity points. For the case of ``const $\triangle z$'', $z_1=0.5$ $z_2=1.0$; for the case of ``const n$\triangle z$'', $z_1=0.2481$ $z_2=0.6787$. Moreover, for the case of ``free $\triangle z$'', different SNIa statistics techniques will give different discontinuity points, which are shown in Table~\ref{frint}.

\begin{figure*}[htbp]
      \centering
      \includegraphics[width=7cm,height=4cm]{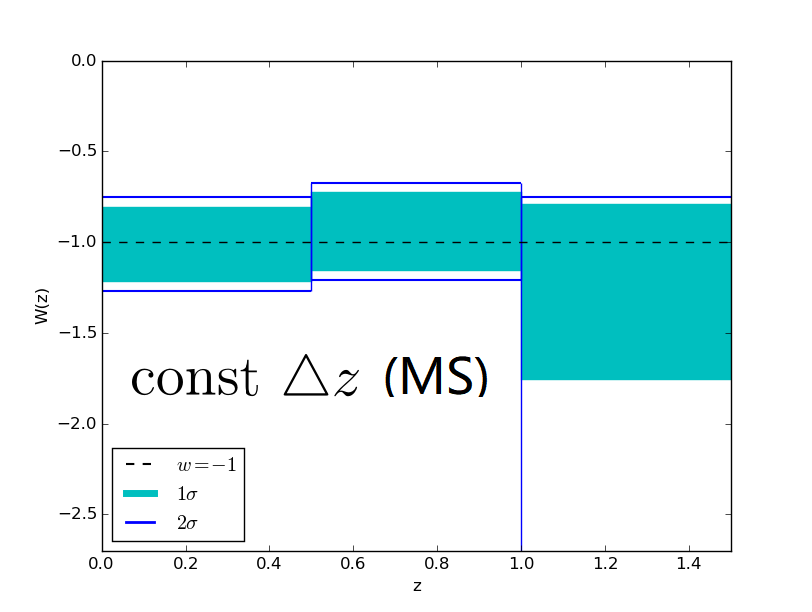}\includegraphics[width=7cm,height=4cm]{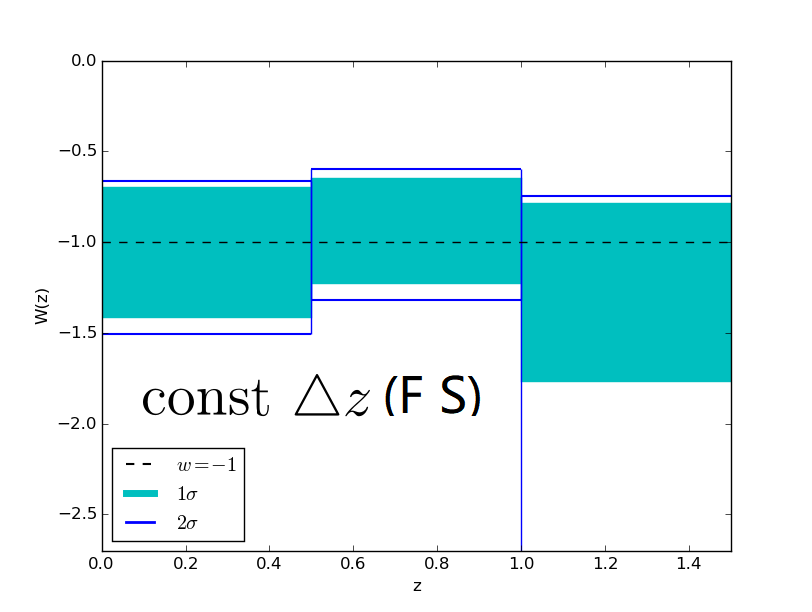}
      \includegraphics[width=7cm,height=4cm]{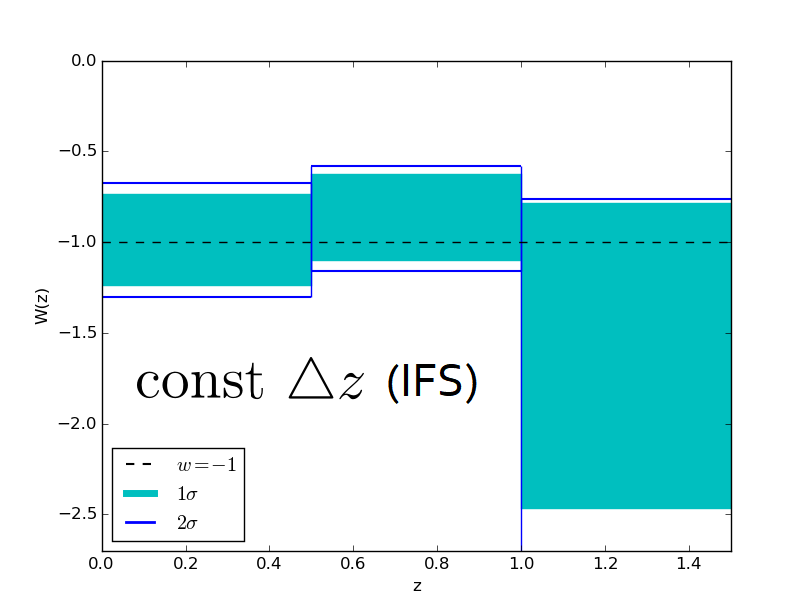}
      \caption{The evolution of w(z) for the ``const $\triangle z$'' case. Three SNIa statistics techniques are used, including MS (upper left panel), FS (upper right panel) and IFS (lower panel). For three different statistics, the 1$\sigma$ and 2$\sigma$ confidence regions have the same evolutionary trend. We find that the largest one is FS and the MS has the minimum confidence regions. In addition, the lines $\omega = -1$ is always contained in the 1$\sigma$ confidence region.  }
      \label{equz}
    \end{figure*}

 In Fig.\ref{equz}, we plot the evolution of $\omega(z)$ for the ``const $\triangle z$'' case. Note that all the three SNIa statistics techniques are taken into account. The cyan rectangular areas of each panel represent the 1$\sigma$ confidence region of $\omega(z)$, the solid blue lines represent the 2$\sigma$ confidence region. In addition, we use dashed black lines to represent the result of $\Lambda CDM$. This figure shows that, different SNIa statistics techniques will give the same evolutionary trend of EoS, i.e., both the first segment and the second segment are smaller than the third one. Moreover, all the results of EoS given by three SNIa statistics techniques are consistent with the prediction of $\Lambda CDM$.

\begin{figure*}[htbp]
      \centering
      \includegraphics[width=7cm,height=4cm]{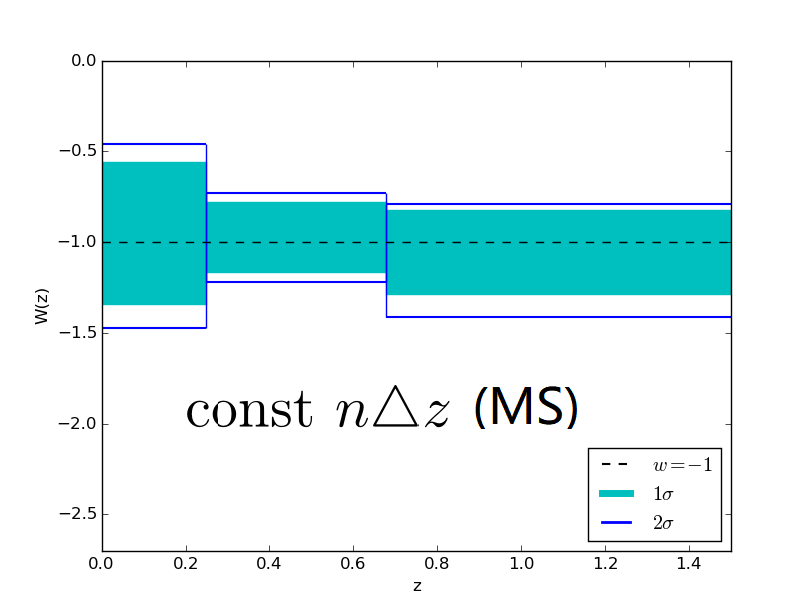}\includegraphics[width=7cm,height=4cm]{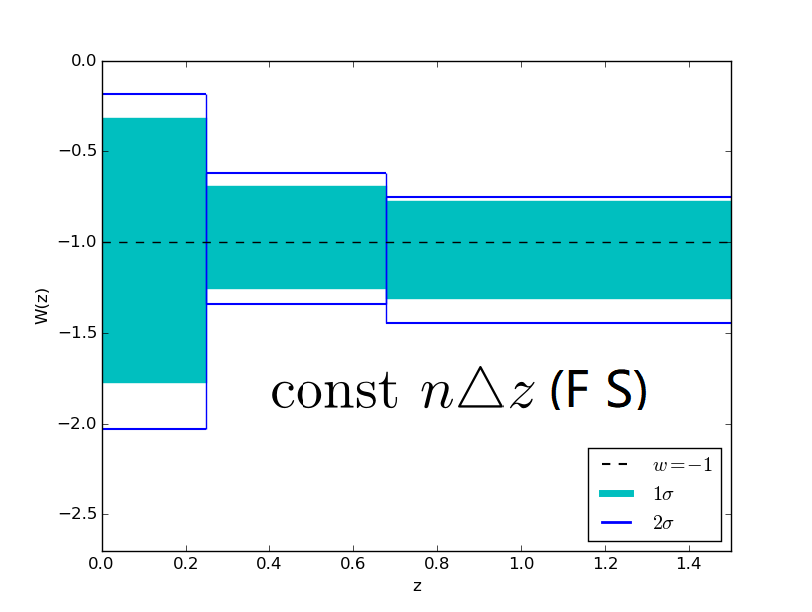}
      \includegraphics[width=7cm,height=4cm]{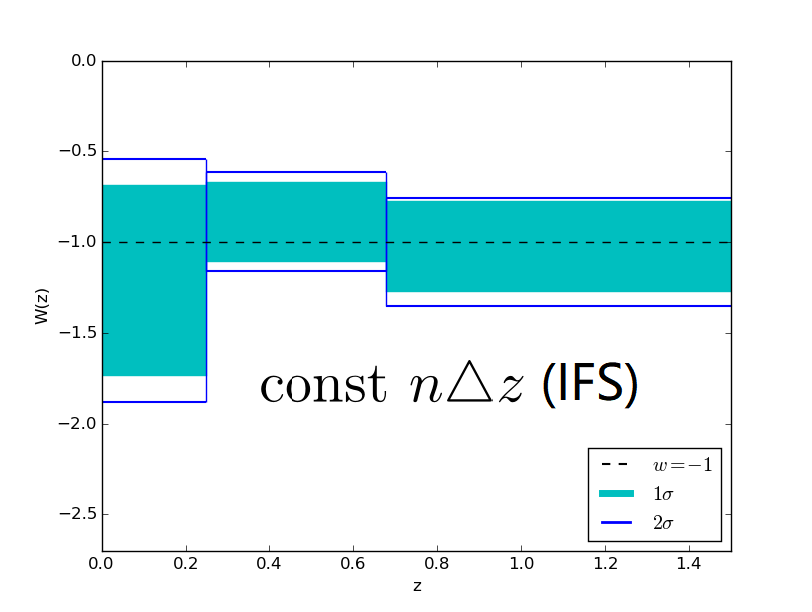}
      \caption{The evolution of w(z) for the case of ``const n$\triangle z$''. Three SNIa statistics techniques are used, including MS (upper left panel), FS(upper right panel) and IFS (lower panel). Comparing to Fig.\ref{equz} , though we adopt different binned methods, the same result can be also concluded that them have the same confidence region size order. Moreover, all the case in this figure contain the EOS $\omega$ of $\Lambda CDM$ in the 1$\sigma$ confidence region.}
      \label{equnz}
    \end{figure*}

 For the case of ``const n$\triangle z$'', we also plot the evolution of $\omega(z)$ in Fig.\ref{equnz}. Comparing to the case of Fig.\ref{equz}, the third segments of $\omega(z)$ have smaller error bars. The reason is that, for the ``const n$\triangle z$'' case, the third redshift interval contains more SNIa samples. Moreover, all the 1$\sigma$ regions of $\omega(z)$ given by three SNIa statistics techniques contain the black dashed lines $\omega = -1$, which correspond to the prediction of $\Lambda CDM$.

\begin{figure*}[htbp]
      \centering
      \includegraphics[width=7cm,height=4cm]{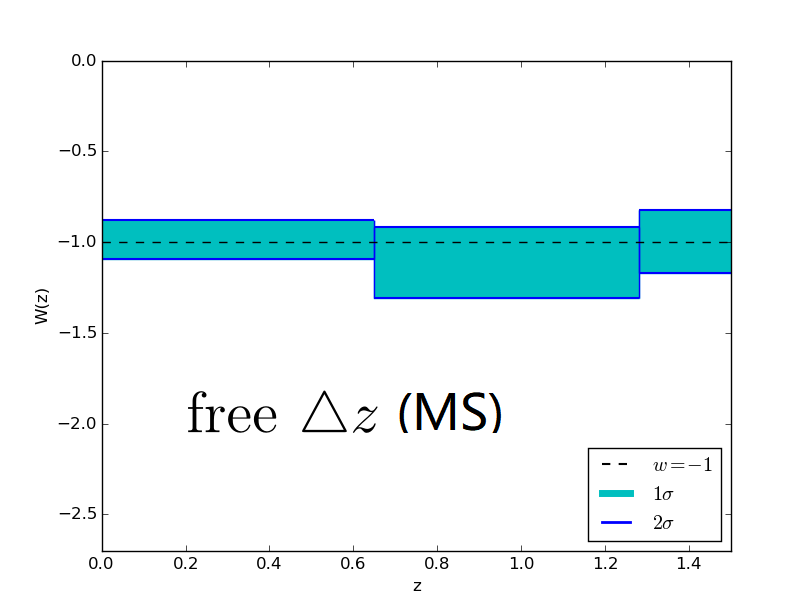}\includegraphics[width=7cm,height=4cm]{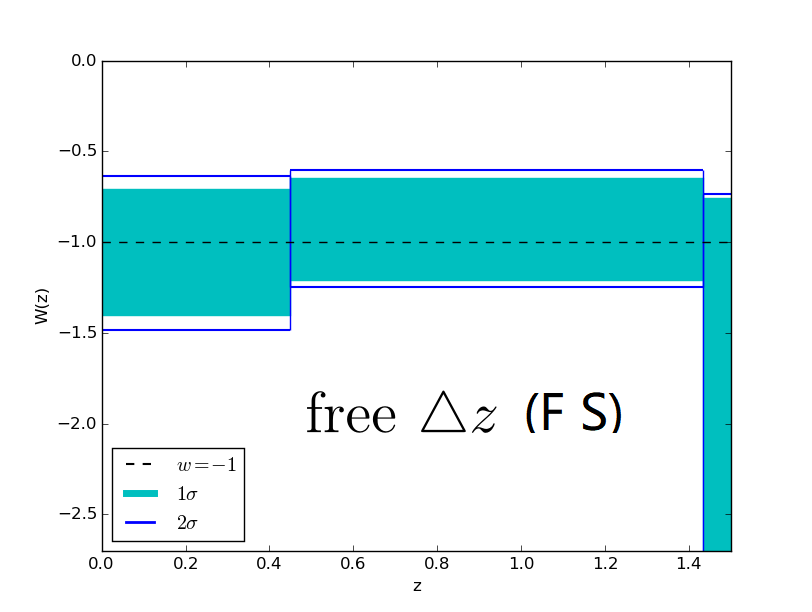}
      \includegraphics[width=7cm,height=4cm]{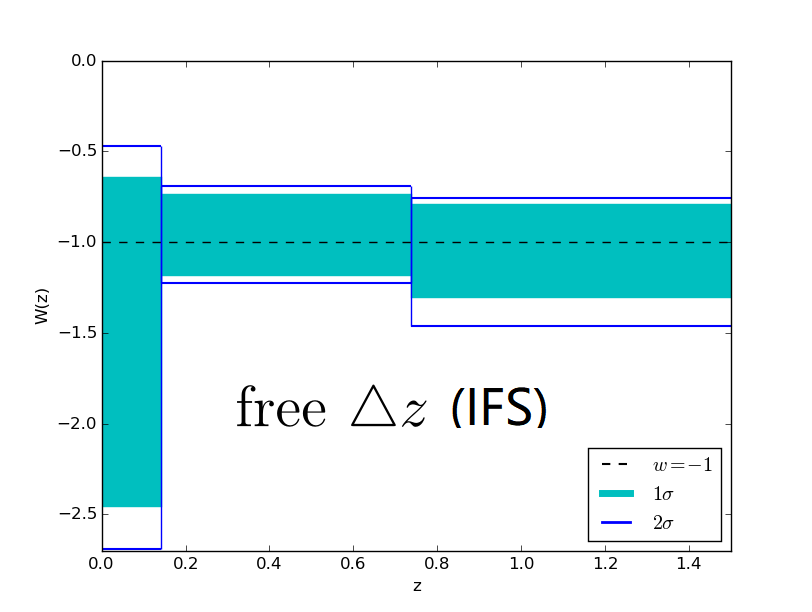}
      \caption{The evolution of w(z) for the case of ``free $\triangle z$''. Three SNIa statistics techniques are used, including MS (upper left panel), FS (upper right panel) and IFS (lower panel). Because of free interval chosen, three different statistics of SINa have different shape of 1$\sigma$ and 2$\sigma$ confidence regions. Even the shape of all confidence regions become different, they are also consistent with the $\Lambda CDM$ like Fig.\ref{equz} and Fig.\ref{equnz}. }
      \label{frez}
    \end{figure*}

 Finally, we discuss the case of ``free $\triangle z$'' model. As shown in Table~\ref{frint}, different model will give different redshift discontinuity points. For this case, the evolution trends of $\omega(z)$ given by different SNIa statistics techniques have significant difference (see Fig.\ref{frez}). However, all the curve of $\omega(z)$ shown in this figure are consistent with the result $\Lambda CDM$.

From Fig.\ref{figcpl} to Fig.\ref{frez}, one can see that all the curves of $\omega(z)$ given by different theoretical models and different SNIa statistics techniques always contain the black dashed line of $\omega =-1$ at the $1\sigma$ confidence region. This proves that $\Lambda CDM$ is favored by current observations.

 \subsection{The effects of adopting different binned parametrizations on parameters estimation}
 In this subsection, we discuss the effects of adopting different binned parametrizations on parameters estimation.

\begin{figure*}[htbp]
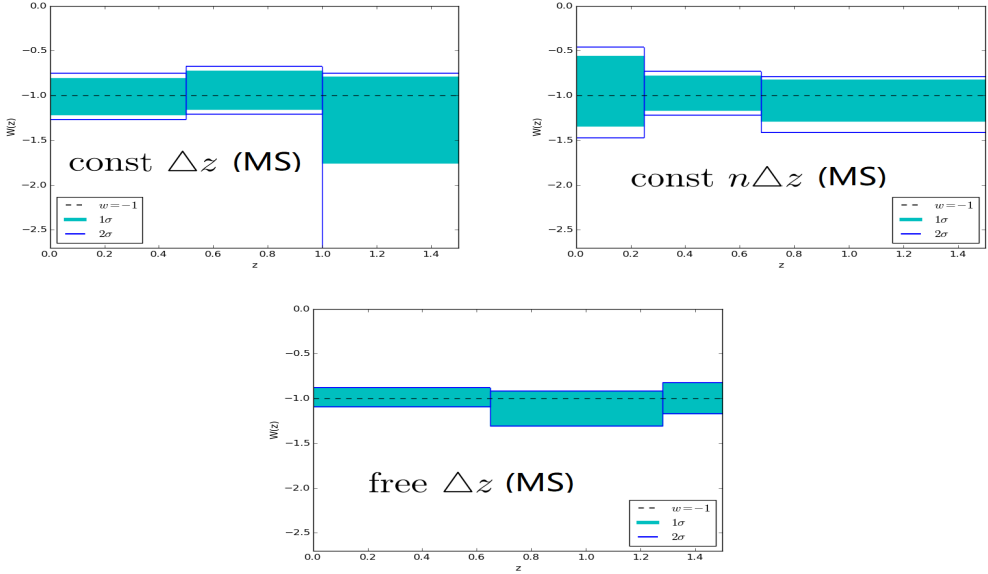

      \centering
      \includegraphics[width=7cm,height=4cm]{Figure2a.png}\includegraphics[width=7cm,height=4cm]{Figure3a.png}
      \includegraphics[width=7cm,height=4cm]{Figure4a.png}
      \caption{The evolution of w(z) for the case of adopting MS. Three binned methods are used, including the const$\triangle z$(upper left panel), the const $n\triangle z$(upper right panel), and the free $\triangle z$(lower panel). Different from the result before, three binned methods have the different evolutionary trend. But it's obvious that the free $\triangle z$ model can find the tightest confidence region. }
      \label{MScompare}
    \end{figure*}

 In Fig.\ref{MScompare}, we plot the evolution curves of $\omega(z)$ for the case of adopting MS. For comparison, three kinds of binned parametrization models are used. It is found that, for each bin, the free $\triangle z$ model always yields the smallest error bars. This conclusion is same as \cite{xiaodongli2011}.

\begin{figure*}[htbp]
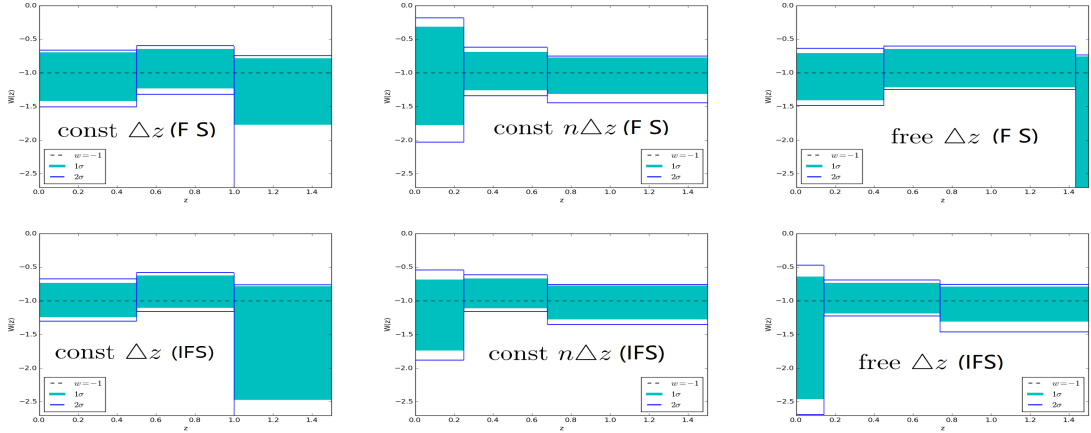

      \centering
      \includegraphics[width=5cm,height=3cm]{Figure2b.png}\includegraphics[width=5cm,height=3cm]{Figure3b.png}
      \includegraphics[width=5cm,height=3cm]{Figure4b.png}
      \includegraphics[width=5cm,height=3cm]{Figure2c.png}\includegraphics[width=5cm,height=3cm]{Figure3c.png}
      \includegraphics[width=5cm,height=3cm]{Figure4c.png}
      \caption{The evolution of w(z) along  for FS (upper panel) and IFS (lower panel). Three binned methods are used, including the const$\triangle z$ (left panel), the const $n\triangle z$ (middle panel), and the free $\triangle z$ (right panel). The free $\triangle z$ model can't find the tightest confidence region obviously. }
      \label{FSIFScompare}
    \end{figure*}

In Fig.\ref{FSIFScompare}, we also plot the evolution curves of $\omega(z)$ for the cases of adopting FS and adopting IFS. It must be emphasized that, in previous literature, the effects of adopting different binned parametrizations have not been discussed in the framework of adopting FS or adopting IFS. From this figure, one can see that the error bars of $\omega(z)$ given by the three binned models are very close. In other words, the conclusion of Fig.\ref{MScompare} does not hold true for the cases of FS and IFS.

  In addition, we also calculate the corresponding FoM in Table~\ref{comparebin}.
  For the case of MS, the free $\triangle z$ model can yield the largest value of FoM, which corresponds to the best observational constrains. This result is consistent with Fig.\ref{MScompare}. However, the free $\triangle z$ model can't yield the largest FoM value for the case of FS and IFS (See the second and third rows in Table~\ref{comparebin}). therefore, we can conclude that the conclusion of Fig.\ref{MScompare} does not hold true for the cases of FS and IFS. It need to be mentioned that, this conclusion has not been obtained in previous literature.

  In summary, it is found that, for the case of MS, the free $\triangle z$ will give the smallest error bars among the three binned parametrizations. However, this conclusion does not hold true for the cases of FS and IFS.

\begin{table*}
\caption{\label{comparebin} The value of FoM for three kinds of binned parametrizations, where all the three SNIa statistics techniques are used}\centering
\begin{tabular*}{\textwidth}{@{}l*{15}{@{\extracolsep{0pt plus12pt}}l}}
\hline\hline
FoM    &const$\triangle z$& const $n\triangle z$  &  free$\triangle z$   \\
\hline
  MS    &1226.42  &1396.20  &2956.18   \\
  FS    &779.39  &678.99  &110.08   \\
  IFS   &780.79 &1123.70  &653.02   \\
\hline
\end{tabular*}
\end{table*}

 \subsection{The effect of adopting different SNIa analysis techniques}
 In this subsection, we discuss the effects of adopting different SNIa statistics techniques on parameters estimation. In order to present the result more clearly, we perform cosmology-fit by only using the SNIa data.
 For simplicity, here we only consider the CPL parametrization and ``const $\triangle z$'' binned parametrization.

 To compare different SNIa statistics techniques, we mainly focus on the parameters estimation of the present fractional density of matter $\Omega_m$, as well as the current deceleration parameter $q_0$, which is given by
 \be
 q_0\equiv \frac{\ddot{a}(t_0)a(t_0)}{\dot{a}(t_0)}=\frac{1}{2}\Omega_m+\Omega_r-\Omega_{de}.
 \ee
 Here $\Omega_r$ and $\Omega_{de}$ are the present fractional densities of radiation and DE, which can be found in Eq.\ref{Fried}. In addition, $t_0$ represents the current time.

\begin{figure*}[htbp]
      \centering
      \includegraphics[width=7cm,height=4cm]{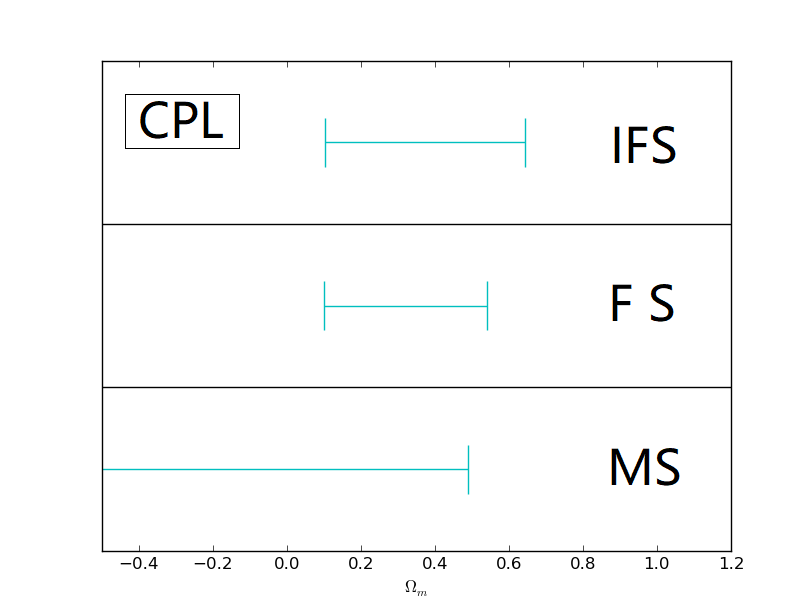}\includegraphics[width=7cm,height=4cm]{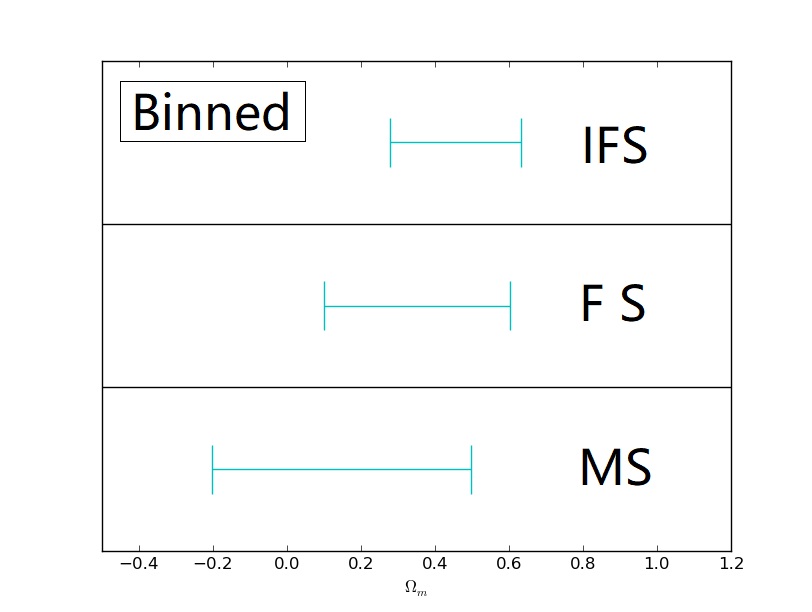}

      \caption{The 1$\sigma$ confidence intervals of $\Omega_m$ for CPL parametrization(left panel) and binned parametrization(right panel). From the bottom to the top of each panel, the three regions represent MS, FS and IFS.}
      \label{onlyom}
    \end{figure*}

 In Fig.\ref{onlyom}, we plot the 1$\sigma$ confidence intervals of $\Omega_m$ for the CPL parametrization and the binned parametrization, where all the three SNIa statistics techniques are. From the bottom to the top of each panel, the three regions represent MS, FS and IFS. For the CPL parametrization, the best-fit results of $\Omega_m$ are $0.023, 0.142$ and $0.481$ for MS, FS and IFS, respectively. For the binned parametrization, $\Omega_m = 0.228, 0.434$ and $0.526$ for MS, FS and IFS, respectively. So we can conclude that IFS yields largest $\Omega_m$. In other words, IFS favor a universe that contains more matter.

\begin{figure*}[htbp]
      \centering
      \includegraphics[width=7cm,height=4cm]{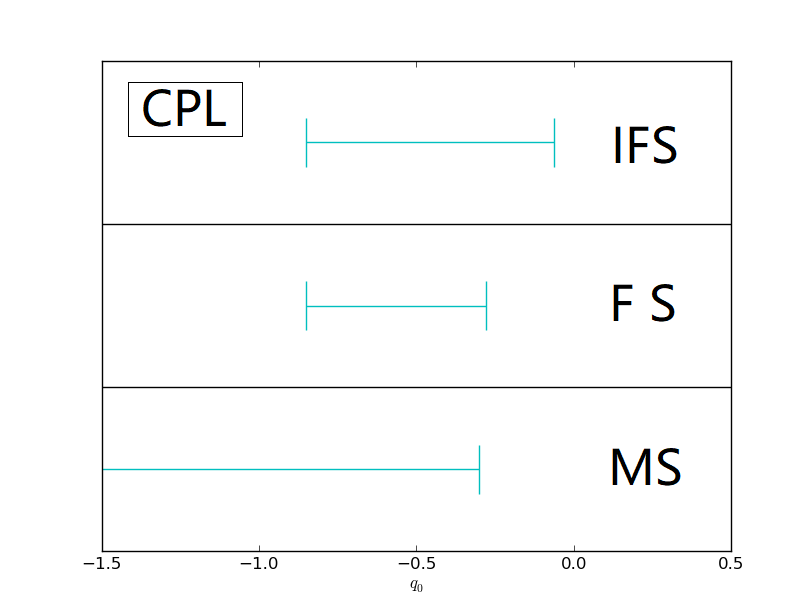}\includegraphics[width=7cm,height=4cm]{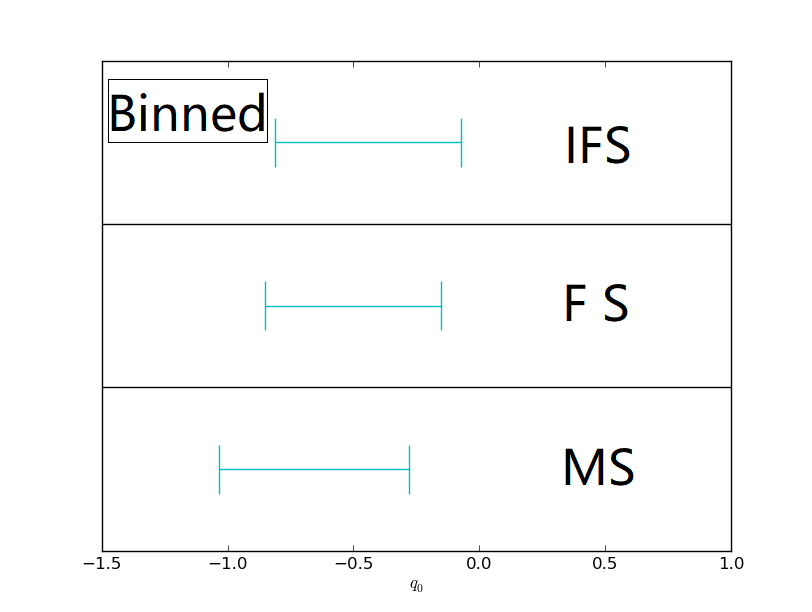}

      \caption{The 1$\sigma$ confidence intervals of current deceleration parameter $q_0$ for CPL parametrization(left panel) and binned parametrization(right panel). From the bottom to the top of each panel, the three regions represent MS, FS and IFS. }
      \label{onlyq}
    \end{figure*}

 In addition, we also present the results of $q_0$ in Fig.\ref{onlyq}. The current deceleration parameter $q_0$ also rapidly varies for different SNIa statistics techniques. For CPL parametrization, the best-fit results of $q_0$ are $-0.966, -0.776$ and $-0.279$ for MS, FS and IFS, respectively. For the binned parametrization, the bset-fit value of $q_0$ are $-0.581, -0.350$ and $-0.211$ for MS, FS and IFS, respectively. One can find that IFS yields largest $q_0$, which reveals a universe with a slowest cosmic acceleration.

 As mentioned above, we conclude that IFS yields larger present fractional density of matter $\Omega_m$ and current deceleration parameter $q_0$. In other words, this SNIa statistics technique favors a universe, which contains more matter and has slower cosmic acceleration. It must be emphasized that this conclusion does not rely on a specific model.

\section{Conclusions and Discussions}\label{conclu}
In our work, we explore the evolution of the DE EOS $\omega$(z) by using model-independent parametriztions. The CPL parametrization and three kinds of binned parametrizations (including ``const$\triangle z$'', ``const $n\triangle z$'' and ``free$\triangle z$'') are taken into account in this work. To perform cosmology-fits, we adopt the observation data including the SNIa observation from JLA samples, the BAO observation from SDSS DR12, and the CMB observation from Planck 2015 distance priors. In particular, for the SNIa data, we make use of three statistics techniques, i.e. MS, FS and IFS.

In previous literature, only one binned parametriztion and one SNIa analysis technique are taken into account.
In this work, we consider all the three binned parametriztions (i.e., const $\triangle z$, const $n\triangle z$ and $free \triangle z$) and all the three SNIa analysis techniques (i.e., MS, FS and IFS). Therefore, we can present a more comprehensive and more systematic study.

Our results are as follows:
\begin{itemize}

\item
For all the cases, $\omega = -1$ is always satisfied at $1\sigma$ confidence regions; It means that $\Lambda$CDM is still favored by current observations ( from Fig.\ref{figcpl} to Fig.\ref{frez}).
\item
For magnitude statistics, ``free $\triangle z$'' model will give the smaller error bars; this conclusion does not hold true for flux statistics and improved flux statistic ( see Fig.\ref{MScompare}, Fig.\ref{FSIFScompare} and Table~\ref{comparebin}).
\item
The improved flux statistic yields a largest present fractional density of matter $\Omega_m$; in addition, this technique will give a largest current deceleration parameter $q_0$ , which reveals a universe with a slowest cosmic acceleration ( see Fig.\ref{onlyom} and Fig.\ref{onlyq}).
\end{itemize}

Recently, the Dark Energy Survey Supernova Program (DES-SN) published the latest SNIa samples.
It would be interesting to analyse the systematic uncertainties of this latest SNIa samples, and then study the corresponding cosmological consequences. This will be done in future works.

\begin{acknowledgments}
 SW is supported by the National Natural Science Foundation of China under Grant No. 11405024 and the Fundamental Research Funds for the Central Universities under Grant No. 16lgpy50.
\end{acknowledgments}


\begin{thebibliography}{}
\bibitem[Ade {et~al.}(2015)]{Planck201514}
 Ade P.A.R., Aghanim N., Arnaud M., {et~al.} 2015, Astron.\ Astrophys, 594, A14.

\bibitem[Alam {et~al.}(2016)]{Alam2016}
 Alam S., Ata M., Bailey S., et al. 2017, \mnras, 470, 2617-2652.

\bibitem[Armendariz-Picon {et~al.}(1999)]{Picon1999}
Armendariz-Picon C., Damour T. and Mukhanov V., 1999, Phys.\ Lett.\ B, 458, 209.

\bibitem[Armendariz-Picon {et~al.}(2001)]{Picon2001}
Armendariz-Picon C., Mukhanov V. and Steinhardt P.J., 2001, \prd, 63, 103510.

\bibitem[Aubourg {et~al.}(2015)]{Aubourg2015}
 Aubourg \`E., Bailey S., Bautista J.E., et al., 2014, \prd, 92, 123516.

\bibitem[Bamba  {et~al.}(2012)]{Bamba2012}
Bamba, K., Capozziello, S., Nojiri, S. and Odintsov, S.D., 2012, Astrophys.\ Space Sci., 342, 155.

\bibitem[Bento {et~al.}(2002)]{Bento2002}
Bento M.C., Bertolami O. and Sen A.A., 2002, \prd, 66, 043507.

\bibitem[Betoule {et~al.}(2014)]{Betoule2014}
 Betoule M., Kessler R., Guy J., {et~al.} 2014, Astron.\ Astrophys, 568, A22.

\bibitem[Cai(2007)]{rgcai2007}
Cai R.G., Phys.\ Lett.\ B, 2007, 657, 228.

\bibitem[Caldwell {et~al.}(1998)]{Caldwell1998}
Caldwell R. R., Dave R. and Steinhardt P.J., 1998, \prl, 80, 1582.

\bibitem[Caldwell(2002)]{Caldwell2002}
Caldwell R. R., 2002, Phys.\ Lett.\ B, 545, 23.

\bibitem[Carroll {et~al}(2003)]{Carroll2003}
Carroll S. M., Hoffman M. and Trodden M., 2003, \prd, 68, 023509.

\bibitem[Chevallier \& Polarski (2001)]{CPL}
Chevallier M. and Polarski D., 2001, Int.\ J.\ Mod.\ Phys.\ D, 10, 213.

\bibitem[Chiba(2000)]{Chiba2000}
Chiba T., Okabe T. and Yamaguchi M., 2000, \prd, 62, 023511.

\bibitem[Einstein(1917)]{Einstein1917}
Einstein, A.~, 1997, Sitzungsber.\ Preuss.\ Akad.\ Wiss.\ Berlin (Math. Phys.), 1917, 142.

\bibitem[Frieman  {et~al.}(2008)]{Frieman2008}
Frieman, J., Turner, M., Huterer, D., 2008, Ann.\ Rev.\ Astron.\ Astrophys., 46, 385.

\bibitem[Gao {et~al.}(2009)]{Gao2009}
Gao C., Wu F. Q., Chen X., et al.  2009, \prd, 79 043511.

\bibitem[Holsclaw {et~al.}(2010)]{Holsclaw2010}
 Holsclaw T., et al. 2010, \prd, 82, 103502.

\bibitem[Huang {et~al.}(2009)]{QGHuang2009}
Huang Q.-G., Li M., Li X.-D. and Wang S., 2009, \prd, 80, 083515.

\bibitem[Huterer \& Starkman(2003)]{Huterer2003}
Huterer D. and Starkman G., 2003, \prl, 90, 031301.

\bibitem[Huterer \& Cooray(2005)]{Huterer2005}
Huterer D. and Cooray A., 2005, \prd, 71, 023506.

\bibitem[Hu \& Sugiyama (1996)]{Hu:1995en}
 Hu W. and Sugiyama N., 1996, \apj, 471, 542.

\bibitem[Hu {et~al.}(2016)]{Hu2016}
Hu Y., { et al.} 2016, Astron.\ Astrophys, 592, A101.

\bibitem[Kamenshchik {et~al.}(2001)]{Kamenshchik2001}
Kamenshchik A.Y., Moschella U. and Pasquier V., 2001 , Phys.\ Lett.\ B, 511, 265.


\bibitem[Lewis\& Bridle(2002)]{Lewis2002}
Lewis A. and Bridle S., 2002, \prd, 66, 103511.

\bibitem[Li(2004)]{hde}
Li M., 2004, Phys.\ Lett.\ B, 603, 1.

\bibitem[Li {et~al.}(2016)]{li2016}
Li M., Li N., Wang S. and Zhou L., 2016, \mnras, 460, 2586.

\bibitem[Li  {et~al.}(2009a)]{MLi2009a}
Li M., Li X.-D., Wang S. et al., 2009, \jcap, 0906, 036.

\bibitem[Li  {et~al.}(2009b)]{MLi2009b}
Li M., Li X.-D., Wang S. et al., 2009, \jcap, 0912, 014.

\bibitem[Li  {et~al.}(2011)]{MLi2011}
Li M., Li X.-D., Wang S. et al., 2011, Commun.\ Theor.\ Phys.\, 56, 525.

\bibitem[Li  {et~al.}(2013)]{MLi2013}
Li M., Li X.-D., Wang S. et al., 2013, Frontiers of Physics, 8, 828.

\bibitem[Linder(2003)]{Linder:2002et}
Linder E.~V., 2003, \prl, 90, 091301.

\bibitem[Li {et~al.}(2011)]{xiaodongli2011}
Li X.-D., Li S., Wang S., Zhang W.-S., Huang Q.-G. and Li M., 2011, \jcap, 1107, 011.

\bibitem[Mohlabeng \& Ralston(2014)]{Mohlabeng2014}
Mohlabeng G.M., Ralston J.P., 2014, \mnras, 439, L16.

\bibitem[Padmanabhan (2003)]{Padmanabhan2003}
Padmanabhan, T.,  2003, \physrep, 235, 380

\bibitem[Peebles \& Ratra(1988)]{Peebles1998}
Peebles P.J.E. and Ratra B., 1988, \apj, 325, L17.

\bibitem[Perlmutter {et~al.}(1999)]{Perlmutter1999}
Perlmutter S., {et al.} 1999, \apj, 517, 565.

\bibitem[Ratra \& Peebles(1988)]{Ratra1998}
Ratra B. and Peebles P.J.E., 1988, \prd, 37, 3406.

\bibitem[Riess {et al.}(1998)]{AGRiess1998}
Riess A.G., et al. 1998, \aj, 116, 1009.

\bibitem[Riess {et~al.}(2007)]{AGRiess2007}
Riess A.G., et al. 2007, \apj, 659, 98.

\bibitem[Shapiro {et~al.}(2004)]{Shapiro2004}
Shapiro S.S., et al. 2004, \prl, 92, 121101.

\bibitem[Shi et al.(2012)]{shi2012}
Shi K., Huang Y. and Lu T., 2012, \mnras, 426, 2452.

\bibitem[Wang {et~al.}(2015)]{WGHZ2015}
Wang S., Geng, J.-J. Hu Y.-L. and Zhang X., 2015, Sci.\ China\ Phys.\ Mech.\ Astron\, 58, 1.

\bibitem[Wang, Li \& Li(2011)]{wll2011}
Wang S., Li X.-D. and Li M., 2011, \prd, 83, 023010.

\bibitem[Wang, Li \& Zhang(2013)]{wlz2013}
Wang S., Li Y.-H and Zhang X., 2013, \prd, 89, 063524

\bibitem[Wang \& Wang(2013)]{WangWang2013}
Wang S. and Wang Y., 2013a, \prd, 88, 043511.

\bibitem[Wang {et~al.}(2014b)]{WWGZ2014}
Wang S., Wang Y.-Z., Geng J.-J. and Zhang X., 2014b, Eur.\ Phys.\ J. C, 74, 3148.

\bibitem[Wang, Wang \& Zhang (2014c)]{WWZ2014}
Wang S., Wang Y.-Z. and Zhang X., 2014c, Commun.\ Theor.\ Phys.\, 62, 927.

\bibitem[Wang, Wang \& Li(2017)]{SWang2017}
Wang S., Wang Y. and Li M., 2017, \physrep, 696, 1-57.

\bibitem[Wang, Wen \& Li(2017)]{Wang2017}
Wang S., Wen S.-X. and Li M., 2017, \jcap, 1703, 037.

\bibitem[Wang, Zhang \& Xia (2008)]{SWang2008}
Wang S., Zhang Y. and Xia T.-Y., 2008, \jcap, 0810, 037.

\bibitem[Wang \& Zhang(2008)]{WZ2008}
Wang S. and Zhang Y., 2008, Phys.\ Lett.\ B, 669 201.

\bibitem[Wang(2000)]{Wang2000}
Wang Y., 2000, \apj, 536, 531.

\bibitem[Wang \& Tegmark(2004)]{wangprl2004}
 Wang Y. and Tegmark M., 2004, \prl, 92, 241302.

\bibitem[Wang \& Mukherjee(2004)]{Wang2004}
 Wang Y. and Mukherjee P., 2004, \apj, 606, 654.

\bibitem[Wang \& Tegmark(2005)]{WangTegmark05}
Wang Y. and Tegmark M., 2005, \prd, 71, 103513.

\bibitem[Wang \& Mukherjee(2007)]{Ywang2007}
 Wang Y. and Mukherjee P., 2007, \prd, 76, 103533.

\bibitem[Wang(2008)]{YunWang2008}
Wang Y., 2008, \prd, 77, 123525.

\bibitem[Wang(2009)]{YunWang2009}
Wang Y., 2009, \prd, 80, 123525.

\bibitem[Wang(2010)]{ywang2010}
Wang Y., 2010, Mod.\ Phys.\ Lett.\ A, 25, 3093.

\bibitem[Wang {et~al.}(2012)]{Wang12CM}
Wang Y., Chuang C.-H. and Mukherjee P., 2012, \prd, 85, 023517.

\bibitem[Wang \& Wang(2013b)]{WangyunWangshuang2013}
Wang Y. and Wang S., 2013b, \prd, 88, 043522.

\bibitem[Wang \& Dai (2016)]{Wang2015}
 Wang Y. and Dai M., 2016, \prd, 94, 08352.


\bibitem[Wei \& Cai(2008)]{weicai2008}
Wei H. and Cai R.G., 2008, Phys.\ Lett.\ B, 660, 113-117.

\bibitem[Wei(2010)]{wei2010}
 Wei H., 2010, Phys.\ Lett.\ B, 687, 286.

\bibitem[Weinberg {et~al.}(2013)]{weinberg2013}
Weinberg D.H., Mortonson M.J., Eisenstein D.J., Hirata C., Riess A.G., Rozo E., 2013, \physrep, 530, 87-255.

\bibitem[Wen, Wang \& Luo(2018)]{Wen2018}
Wen S.-X., Wang S. and Luo X.-L., 2018, \jcap, 1807, 011.

\bibitem[Wetterich(1988)]{Wetterich1988}
Wetterich C., 1988, Nucl.\ Phys.\ B, 302, 668.

\bibitem[Xia \& Zhang(2007)]{xiazhang2007}
Xia T.Y. and Zhang Y., 2007, Phys.\ Lett.\ B, 656, 19.

\bibitem[Zhang {et~al.}(2006)]{Zhang2006}
Zhang X., Wu F.Q. and Zhang J., 2006, \jcap, 0601, 003.

\bibitem[Zhang {et~al.}(2007)]{YZhang2007}
Zhang Y., Xia T.Y. and Zhao W., 2007, Class.\ Quant.\ Grav.\, 24, 3309.

\bibitem[Zlatev(1999) {et~al.}]{Zlatev1999}
Zlatev I., Wang L. and Steinhardt P.J., 1999, \prl, 82, 896.


\end{thebibliography}
\end{document}